\documentclass[aps,amssymb,noshowpacs,11pt,eqsecnum,nofootinbib]{revtex4}

\usepackage{graphicx,graphics}
\usepackage{enumerate}
\usepackage{subfigure}
\usepackage{amsmath}
\usepackage{amsfonts}
\usepackage{amssymb}
\usepackage{amsthm}

\begin{document}

%\title{Exact Constants for Spanning Trees, Loop Erased Random Walks and Abelian Sandpile}
\title{Return probability for the loop-erased random walk and\\
mean height in sandpile : a proof}

\author{V.S. Poghosyan$^{1}$, V.B. Priezzhev$^2$ and P. Ruelle$^1$}
\affiliation{
$^1$Institut de Recherche en Math\'ematique et Physique,
Universit\'{e} catholique de Louvain, B-1348 Louvain-La-Neuve, Belgium\\
$^2$Bogoliubov Laboratory of Theoretical Physics,
Joint Institute for Nuclear Research, 141980 Dubna, Russia
}
\begin{abstract}
Single site height probabilities in the Abelian sandpile model, and the corresponding mean height $\langle h \rangle$, are directly related to the probability $P_{\rm ret}$ that a loop erased random walk passes through a nearest neighbour of the starting site (return probability). The exact values of these quantities on the square lattice have been conjectured, in particular $\langle h \rangle = 25/8$ and $P_{\rm ret} = 5/16$. We provide a rigourous proof of this conjecture by using a {\it local} monomer-dimer formulation of these questions. 
%For the loop erased random walk on the large square lattice, we prove that the return probability
%to the neighboring lattice site is 5/16. For the Abelian sandpile model, we prove that the average height
%in the recurrent state is 25/8.
\end{abstract}
%\pacs{05.40.Fb, 02.10.Ox, }

\maketitle

%\noindent \emph{Keywords}:  Abelian sandpile model, loop-erased random walk, monomer-dimer model.

\section{Introduction}

As different as they may appear at first sight, the dimer model, the loop-erased random walk (LERW) and the Abelian sandpile model (ASM) are very closely connected, as they all have an alternative formulation in terms of a unifying concept: spanning trees. 

Among the three models just mentioned, the dimer model is the oldest one as it was formulated in 1937 by  Fowler and Rushbrooke \cite{foru}, although the first exact results were obtained somewhat later, in the sixties, by Kasteleyn, Fisher, Temperley and Stephenson \cite{kast1,kast2,fish,fite,fist}. The dimer (or domino)  model, also known as the perfect matching problem, has been the subject of an increasing number of works over the last decade, see for instance \cite{ken} and the references therein. The correspondence between packed dimer configurations (no vacancy or monomer) on the square lattice and spanning trees on a sublattice was established by Temperley \cite{temp}, and generalized to general planar graphs by Burton and Pemantle \cite{bupe}. This correspondence holds in the presence of monomers, and leads to spanning webs rather than spanning trees \cite{bbgj,ppr}. It will be recalled in Section 3.

The loop-erased random walk was defined by Lawler \cite{law} as paths generated from a simple symmetric random walk by removing the loops as they appear. The connection between the LERW and the spanning trees has been discussed in several works \cite{br,al,pe,Majumdar,wil}, with the result that the probability measure on LERW sample paths coincides with the uniform measure on chemical paths of spanning trees. 

Finally the Abelian sandpile model is an open stochastic dynamical system, which works like a non-linear diffusion process \cite{btw,dhar}. The ASM is defined in terms of height variables, attached to the sites of a two-dimensional square lattice. The height variables $h_i$ take the four integer values 1, 2, 3 and 4, but only a fraction of all height configurations, called recurrent, keep reoccuring when the dynamics is run over long periods. The long time behaviour of the model is controlled by the stationary measure on the height configurations. This measure is uniform on the recurrent configurations and vanishes on the non-recurrent ones. The connection to spanning trees stems from the burning algorithm \cite{madh1}, which establishes a direct and one-to-one correspondence between the recurrent configurations of the sandpile on a lattice and spanning trees on the same lattice.

All three models have non-local features. This is manifest for the dimer model and the LERW, whereas the non-locality in the ASM comes from the recurrence criterion \cite{dhar}, which requires to scan the whole of a height configuration before one can declare it recurrent. These non-local features usually make explicit calculations particularly hard. Let us also mention that the question of conformal invariance has been addressed in these three models, directly in terms of a conformal field theory for the dimer model \cite{kenyon,iprh,plf}, spanning webs \cite{bgpt} and the ASM \cite{maru,ru,jpr,pgpr}, and as stochastic Loewner processes (SLE) in the case of LERW and spanning trees \cite{schr,lasw}.

The purpose of this work is to compute the value of the return probability for the LERW $P_{\rm ret} = {5 \over 16}$, announced in \cite{popr}. This value was based on the observation that the return probability and the ASM mean height are related by $P_{\rm ret} = {\langle h \rangle \over 2}-{5 \over 4}$, and on an earlier conjecture made in \cite{jpr} for the mean height $\langle h \rangle = {25 \over 8}$, in the limit of the infinite square lattice. It turns out that these two numbers are themselves related to three others, recently introduced by Levine and Peres \cite{lepe}, namely the looping constant $\xi$, the ratio $\tau$ between the number of spanning unicycles and the number of spanning trees, and the mean length $\lambda$ of the cycle in a spanning unicycle, all defined as limits over increasing finite square grids. In addition these five numbers have all a $d$-dimensional analogue, and remain rationally related in any dimension \cite{lepe}. In this sense, our result provides a proof for the values of these five numbers in two dimensions, $P_{\rm ret}={5 \over 16}$, $\langle h \rangle = {25 \over 8}$, $\xi = {5 \over 4}$, $\tau={1 \over 8}$ and $\lambda=8$.

In Section 2, we recall the expressions of the LERW return probability and of the ASM mean height in terms of spanning trees with some specific properties. Section 3 contains the proof itself and reduces the counting of the required spanning trees to certain local configurations of monomers and dimers. We should emphasize that the proof does not rely on the exact evaluation of the multiple integral on which the conjecture made in \cite{jpr} is based. On the contrary the proof is essentially combinatorial, and shows that the counting of spanning graphs with certain non-local properties can be reduced to the counting of local monomer-dimer arrangements, which can then be easily carried out. It thus avoids the full complexity of the graph theoretical computations, inherent in \cite{priez,jpr}. Our proof shows and explains why the LERW return probability and the other four related quantities are such simple numbers.

%%%%%%%%%%%%%%%%%%%%%%%%%%%%%%%%%%%%%%%%%%%%%%%%%%%%%%%%%%%%%%%%%%%%%%%%%%%%%%%%%%%%%%

\section{The LERW return probability and the ASM mean height}

The Abelian sandpile model \cite{btw} in finite volume is defined by height variables, located at the sites of a finite, two-dimensional square grid and taking the values $1,2,3,4$ in stable configurations. Particles are added one by one at a random site, thereby increasing by 1 the value of the height at that site. If the height exceeds $4$, then the site becomes unstable and topples, transferring one particle to each of its neighbouring sites (sand may fall off the system at boundaries). As a consequence, one (or more) neighbour may become unstable, in which case it topples too, and so on for the neighbours of the neighbours. When no unstable site remains, another particle is dropped at a random site and the relaxation process repeated. 

The analysis of this discrete dynamics was performed by Dhar \cite{dhar}, who introduced the notion of recurrent configuration. He showed that the system enters the subset of recurrent configurations after a finite time, which depends on the initial configuration, and never leaves it. Therefore the probability distribution which controls the asymptotic behaviour of the sandpile vanishes on non-recurrent configurations; it can be shown to be uniform on the recurrent subset (of size $\sim 3.21^{L^2}$ for a $L \times L$ grid). Dhar also gave a criterion to select the recurrent configurations. Its explicit form will not be important for what follows.

As sandpile configurations are made of random variables valued in $\{1,2,3,4\}$ and distributed according to the uniform measure on the recurrent subset, it is natural to ask about the statistical properties of these variables and their spatial correlations. In particular the first question concerns the distribution of the height at a single site. As a first step, and in order to avoid boundary effects, we consider a site deep in the middle of the grid, and take the infinite volume limit. We denote by $P_i$, for $i=1,2,3,4$, the resulting probabilities, namely $P_i$ is the probability that the height at any fixed site be equal to $i$, in the limit of an infinite grid.

The first of these four probabilities has been obtained by Majumdar and Dhar \cite{madh2},
\begin{equation}
P_1=\frac{2}{\pi^2}-\frac{4}{\pi^3} \simeq 0.07363.
\label{P1}
\end{equation}
The other three probabilities $P_2$, $P_3$ and $P_4$ turned out to be more complicated, and are more conveniently expressed in terms of spanning trees. As mentioned in the Introduction, the burning algorithm \cite{madh1}, which is the algorithmic translation of the recurrence criterion found by Dhar \cite{dhar}, establishes a one-to-one mapping between the set of recurrent configurations and the set of rooted (oriented) spanning trees on the same lattice. The characterization of those spanning trees which correspond, under this mapping, to recurrent configurations with certain height values at certain positions, has been worked out in \cite{priez}. As a result, the fractions $P_i$ of recurrent configurations which have a height equal to $i$ at a reference site, are given by the following fractions among spanning trees,
\begin{equation}
P_1 = \frac{X_0}{4\, \mathcal{N}}\;,\quad
P_2 = P_1 + \frac{X_1}{3\, \mathcal{N}}\;,\quad
P_3 = P_2 + \frac{X_2}{2\, \mathcal{N}}\;,\quad
P_4 = P_3 + \frac{X_3}{    \mathcal{N}}\;,
\label{siteprob}
\end{equation}
where $X_k$, for $k=0,1,2,3$, is the number of spanning trees such that the reference site has
exactly $k$ predecessors among its four nearest neighbours \cite{priez} (a site $x$ is called a predecessor of $y$ if the path along the tree from $x$ to the root passes through $y$). 
The quantity $\mathcal{N}$ is the total number of spanning trees on the grid. The identities (\ref{siteprob}) are valid for a finite grid. Their infinite volume limits exist, and define the sought probabilities.

Unlike $X_0$, the quantities $X_k$ for $k \geq 1$ are more complicated to compute because one has to count the spanning trees satisfying a non-local constraint (those in $X_0$ satisfy local constraints). These numbers have been first computed in \cite{priez}, the results taking the following form,
\begin{eqnarray}
&&P_2=\frac{1}{2}-\frac{3}{2\pi}-\frac{2}{\pi^2}+\frac{12}{\pi^3}+\frac{I_1}{4}, \label{PP2} \\
&&P_3=\frac{1}{4}+\frac{3}{2\pi}+\frac{1}{\pi^2}-\frac{12}{\pi^3}-\frac{I_1}{2}-\frac{3I_2}{32}, \label{PP3} \\
&&P_4=\frac{1}{4}-\frac{1}{\pi^2}+\frac{4}{\pi^3}+\frac{I_1}{4}+\frac{3I_2}{32}, \label{PP4}
\end{eqnarray}
where $I_1$ and $I_2$ are two complicated, multiple integrals. Their numerical evaluation gave the following values for the probabilities, $P_2 \simeq 0.1739$, $P_3 \simeq 0.3063$ and $P_4 \simeq 0.4461$.

%Here $I_{\nu}$, $\nu=1,2$ are integrals
%\begin{equation}
%I_{\nu}=\frac{1}{(2\pi)^4}\int\!\!\!\!\!\int\!\!\!\!\!\int\!\!\!\!\!\int_{0}^{2\pi}\!\!\!\!\!
%\frac{i \sin \beta_1 \det M_{\nu} \, d\alpha_1 d\alpha_2 d\beta_1 d\beta_2}
%{D(\alpha_1,\beta_1)D(\alpha_2,\beta_2)D(\alpha_1+\alpha_2,\beta_1+\beta_2)},
%\label{I}
%\end{equation}
%where
%\begin{equation}
%D(\alpha,\beta)=2-\cos(\alpha)-\cos(\beta)
%\label{D}
%\end{equation}
%and $ M_1,M_2 $ are matrices
%\begin{equation}
%M_1=
%\begin{pmatrix}
%1 & 1 & e^{i\alpha_2} & 1 \\
%3 & e^{i(\beta_1+\beta_2)} & e^{i(\alpha_2-\beta_2)} & e^{-i\beta_1} \\
%4/\pi-1 & e^{i(\alpha_1+\alpha_2)} & 1 & e^{-i\alpha_1} \\
%4/\pi-1 & e^{-i(\alpha_1+\alpha_2)} & e^{2i\alpha_2} & e^{i\alpha_1}
%\end{pmatrix},
%\label{M1}
%\end{equation}
%\begin{equation}
%M_2=
%\begin{pmatrix}
%e^{i\beta_2} & e^{-i(\alpha_1+\alpha_2)-i(\beta_1+\beta_2)} & e^{i\beta_1} \\
%e^{-i\alpha_2} & 1 & e^{-i\alpha_1} \\
%e^{i\alpha_2} & e^{-2i(\alpha_1+\alpha_2)} & e^{i\alpha_1}
%\end{pmatrix}.
%\label{M2}
%\end{equation}

Somewhat later, the calculation of $P_i$ on the plane (and the upper half-plane) was reconsidered in \cite{jpr}, which led to an exact relation between $P_2$ and $P_3$,
\begin{equation}
(\pi-8)P_2+2(\pi-2)P_3=\pi-2-\frac{3}{\pi}+\frac{12}{\pi^2}-\frac{48}{\pi^3},
\end{equation}
and a similar relation between the two integrals $I_1$ and $I_2$. A conjecture on the value of the remaining integral, based on its numerical evaluation to twelve decimal places (later pushed to 25 places), then yielded the following conjectural values for the probabilities \cite{jpr},
\begin{equation}
P_2 = \frac{1}{4} - \frac{1}{2\pi} - \frac{3}{\pi^2} + \frac{12}{\pi^3}\,,\quad
P_3 = \frac{3}{8} + \frac{1}{\pi} - \frac{12}{\pi^3}\,,\quad
P_4 = \frac{3}{8} - \frac{1}{2 \pi} + \frac{1}{\pi^2} + \frac{4}{\pi^3}\,,
\end{equation}
and for the stationary mean height (particle density),
\begin{equation}
\langle h \rangle = P_1 + 2P_2 + 3P_3 + 4P_4 = \frac{25}{8}.
\end{equation}
Conjectured by Grassberger \cite{DharGrass} almost 20 years ago, the value $25/8$ has remained an enigmatic fact of the Abelian sandpile for its striking simplicity. We will prove this conjecture, not by computing the mean height within the sandpile model, but rather by using a recent observation concerning the LERW return probability. 

For a LERW starting from the origin, we let the return probability $P_{\rm ret}$ be the probability that a path passes through a fixed nearest neighbour of the origin, say the right neighbour. In terms of uniformly distributed spanning trees, $P_{\rm ret}$ is the probability that the origin is a predecessor of its right neighbouring site. By listing explicitly the situations where this is the case among those where the right neighbour has $k$ predecessors among its nearest neighbours, the following identity follows \cite{popr}
\begin{equation}
P_{\rm ret} = \frac{X_1}{4\mathcal{N}} + \frac{X_2}{2\mathcal{N}} + \frac{3 X_3}{4\mathcal{N}}.
\end{equation}

By using the equations (\ref{siteprob}) to trade the $X_k$ for the $P_i$ (and the identity $P_1 + P_2 + P_3 + P_4 = 1$), the previous combination turns out to be directly related to the mean height of the sandpile model, 
\begin{equation}
P_{\rm ret} = \frac{1}{4} (- 3 P_1 - P_2 + P_3 + 3 P_4) = \frac{1}{2} (P_1 + 2 P_2 + 3 P_3 + 4 P_4) - \frac{5}{4} = \frac{\langle h \rangle}{2} - \frac{5}{4}.
\end{equation}

\begin{figure}[h]
\includegraphics[width=135mm]{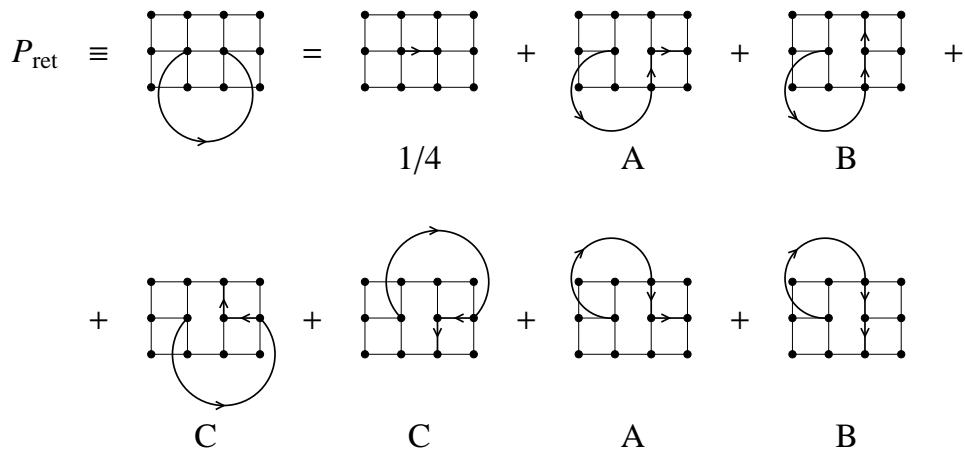}
\caption{\label{pred1} Decomposition of th return probability in terms of chemical paths between two nearest neighbours.}
\end{figure}

In order to actually compute the return probability, we express directly the fact that the origin is a predecessor of its right neighbour, without paying attention to the other nearest neighbours. This is pictured in Fig.\,\ref{pred1}, where all possible configurations are shown.

If $A,B,C$ denote the corresponding fractions of spanning trees with the chemical paths as shown, we have that 
\begin{equation}
P_{\rm ret} = \frac{1}{4} + 2 \, (A + B + C).
\label{predprob}
\end{equation}
It remains to compute $A,B$ and $C$. This will be done in the next section by using particular local arrangements of monomers and dimers.

%%%%%%%%%%%%%%%%%%%%%%%%%%%%%%%%%%%%%%%%%%%%%%%%%%%%%%%%%%%%%%%%%%%%%%%%%%%%%%%%%%%%%%

\section{Monomer-dimer computation of LERW return probability}

Consider a $(2n-1)\times (2n-1)$ square lattice $\mathcal{L}$ with the rightmost lower site removed. The sites of the lattice can be subdivided into three subsets: (1) black sites forming the sublattice $\mathcal{B}$ of sites with odd-odd coordinates; (2) white sites forming the sublattice $\mathcal{W}$ of sites with even-even coordinates; (3) the other remaining sites colored grey. The corner site removed is black and denoted below by $r$ (for root).

\begin{figure}[!ht]
\includegraphics[width=150mm]{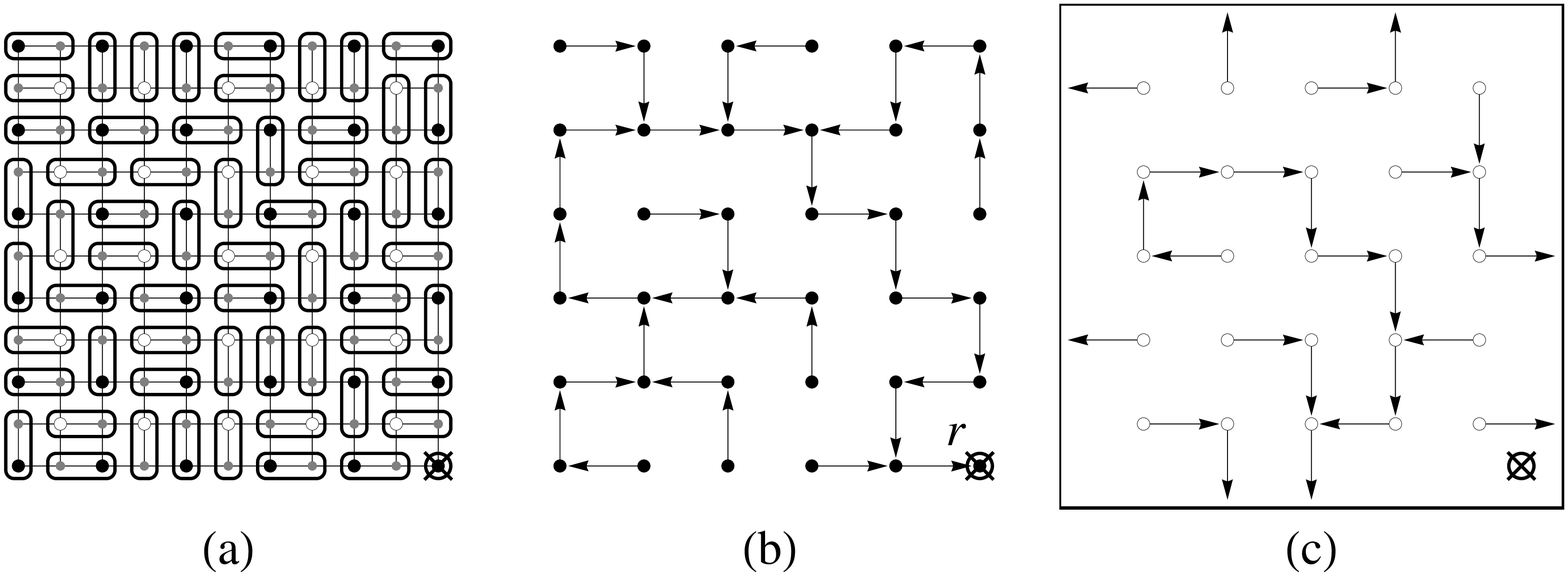}
\caption{\label{dim1} (a) Dimer tiling; (b) spanning tree on the odd-odd sublattice or (c) even-even sublattice.}
\end{figure}

Consider a dense packed dimer configuration on $\mathcal{L}$. Each dimer covers two sites, of colors black and grey, or white and grey. Temperley's correspondence associates two sets of arrows to the dimer configuration in the following way. We replace each dimer covering a black site by an arrow directed from the black site to the grey one, and similarly we replace each dimer covering a white site by an arrow directed from the white site to the grey one. The sets of black and white arrows are two acyclic configurations of arrows, which form spanning trees on the two sublattices $\mathcal{B}$ and $\mathcal{W}$, see Fig.\,\ref{dim1}. The two spanning trees are dual, and from either one, the original dimer configuration can be entirely reconstructed.

A directed path from a site $x \in \mathcal{B}$ ($x \in \mathcal{W}$) to a site $y \in \mathcal{B}$ ($y \in \mathcal{W}$) along the branches of the black (white) spanning tree is the chemical path from $x$ to $y$. As shown in \cite{br,al,pe,Majumdar,wil}, the statistical properties of the chemical paths coincide with those of the loop erased random walk. Our main interest here is the return probability of the LERW: the probability that a chemical path starting at a given point on the spanning tree visits one of neighbouring sites of this point, say the right one, before going off to the root located at $r$.

\begin{figure}[t]
\includegraphics[width=150mm]{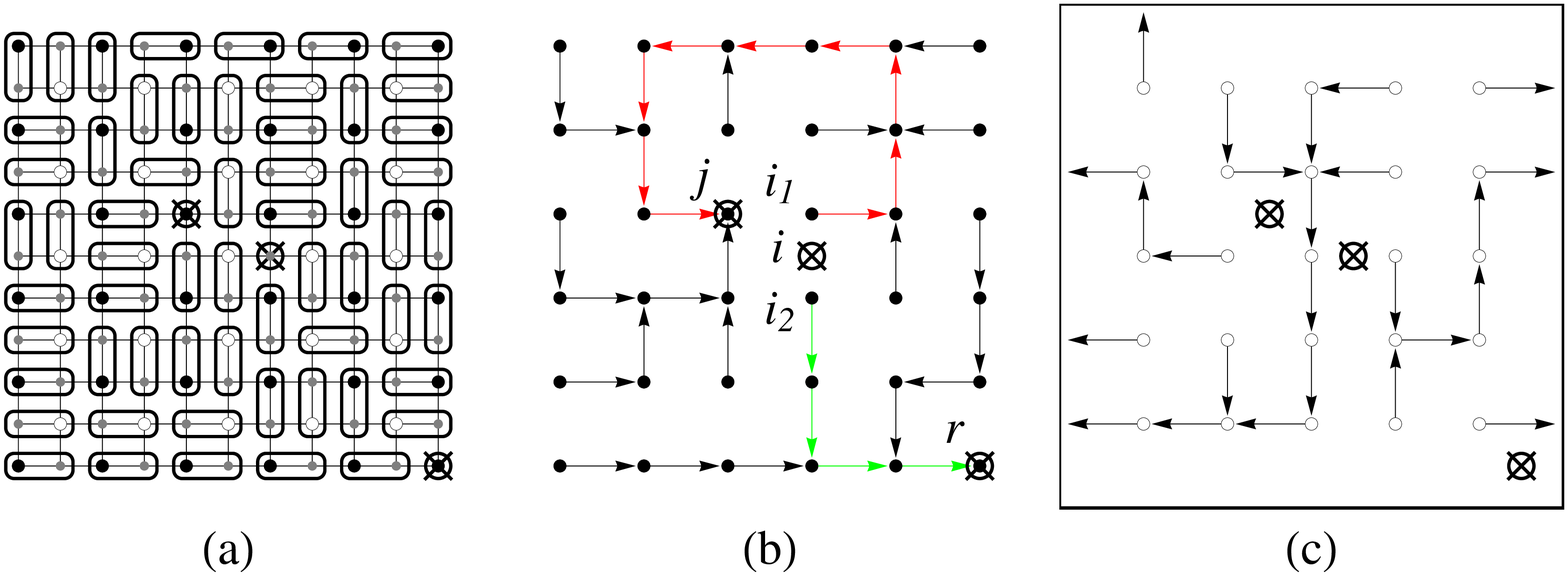}
\vspace{-5mm}
\caption{\label{dim2} (a) Dimer tiling with two monomers inside the lattice; (b) and (c) spanning trees corresponding to case (I). }
\end{figure}

We consider now packed dimer tilings on the lattice $\mathcal{L}'$ obtained from the $(2n-1)\times (2n-1)$ lattice by removing three sites: the corner $r$, a grey site $i$ and a black site $j$. The grey site $i$ has two nearest neighbors $i_1, i_2 \in \mathcal{B}$, see Fig.\ref{dim2}. Kenyon made the observation (lemma 17 in \cite{Kenyon}) that if $j$ was on the boundary of $\mathcal{L}'$, then the sites $i_1$ and $i_2$ would have to be in different components of the two-component black spanning tree, one component being rooted at $r$, the other at $j$. Indeed, $i_1$ and $i_2$ cannot be in the same component for otherwise the chemical path from $i_1$ to $i_2$ together with the bond of $\cal B$ linking $i_1$ and $i_2$ would form a loop enclosing an odd number of lattice sites of $\mathcal{L}'$ which cannot be fully covered by dimers. When $j$ is not on the boundary, this result does not hold because one of the paths from $i_1$ or from $i_2$ may form a loop around $j$ (so that one of $i_1,i_2$ is directed to $r$, the other to a loop winding around $j$), or else both $i_1$ and $i_2$ are directed to $r$, the chemical path between the two going around $j$.

For our purposes, we will choose the site $j$ as the left nearest neighbour of $i_1$ on the sublattice $\mathcal{B}$. Then a closed loop around $j$ is impossible for the lack of space between $j$ and $i_1$ or $i_2$ on the sublattice $\mathcal{B}$. Three possibilities remain.

\begin{figure}[!ht]
\includegraphics[width=130mm]{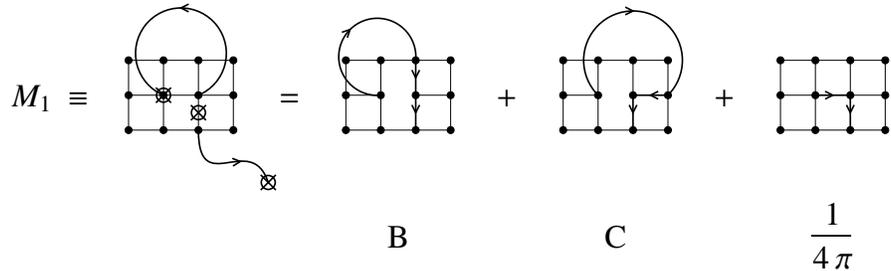}
%\vspace{-5mm}
\caption{\label{pred3} The tree possible path configurations for case (I).}
\end{figure}

(I) The first possibility is depicted in Fig.\,\ref{dim2}b: the path from $i_1$ goes to $j$ and the path from $i_2$ goes to $r$. By reversing the orientation of the first (red) path and inserting an arrow from $i_1$ to $i_2$, we obtain a long path from $j$ to $r$ passing through the bond $(i_1,i_2)$.
Depending on the bonds used by this path around $i_1$, we have three possible configurations shown in Fig.\,\ref{pred3}. 

The first two are exactly those we had denoted by $B$ and $C$ in the previous section (up to a mirror transformation). The third one corresponds to all spanning trees on $\cal B$ which use the two bonds $(j,i_1)$ and $(i_1,i_2)$. In terms of dimers on the original lattice ${\cal L}$, it is the set of all dimer coverings with two forced dimers on $j$ and $i_1$. So the fraction of these is a local dimer-dimer correlation, computed by Fisher and Stephenson in \cite{fist}, and equal to $1/4\pi$ in the limit of large lattices.

\begin{figure}[!ht]
\includegraphics[width=150mm]{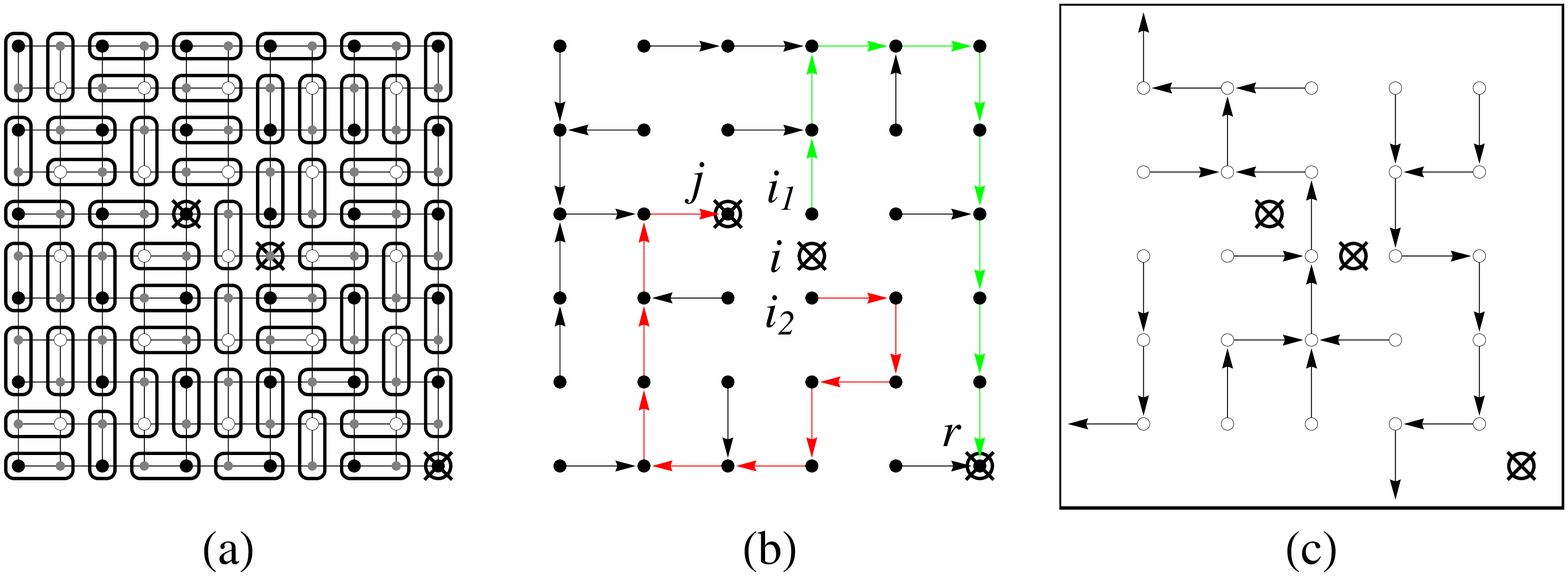}
\vspace{-5mm}
\caption{\label{dim3} The tiling and spanning trees for the case (II).}
\includegraphics[width=130mm]{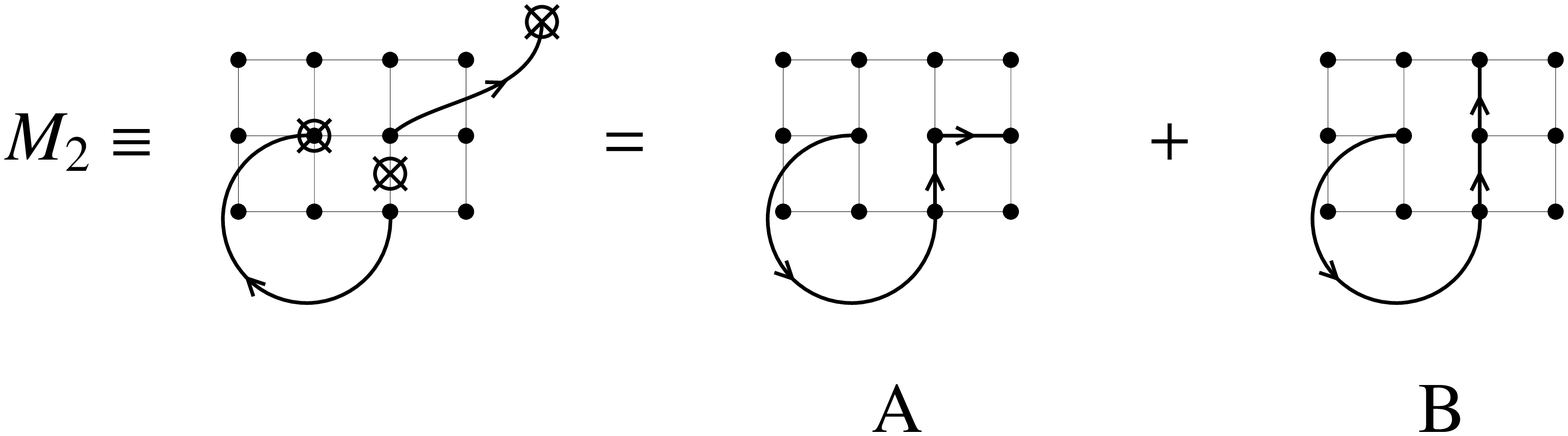}
\vspace{-8mm}
\caption{\label{pred4} The two possible path configurations for the case (II).}
\end{figure}

(II) The path from $i_1$ goes to $r$ and the path from $i_2$ goes to $j$, see Fig.\,\ref{dim3}b. Like in case (I), we reverse the orientation the second path and insert an arrow form $i_2$ to $i_1$, to obtain a long path from $j$ to $r$ passing through the bond $(i_2,i_1)$. It leads to the two possible local configurations $A$ and $B$ shown in Fig.\,\ref{pred4}.

\begin{figure}[!ht]
\includegraphics[width=150mm]{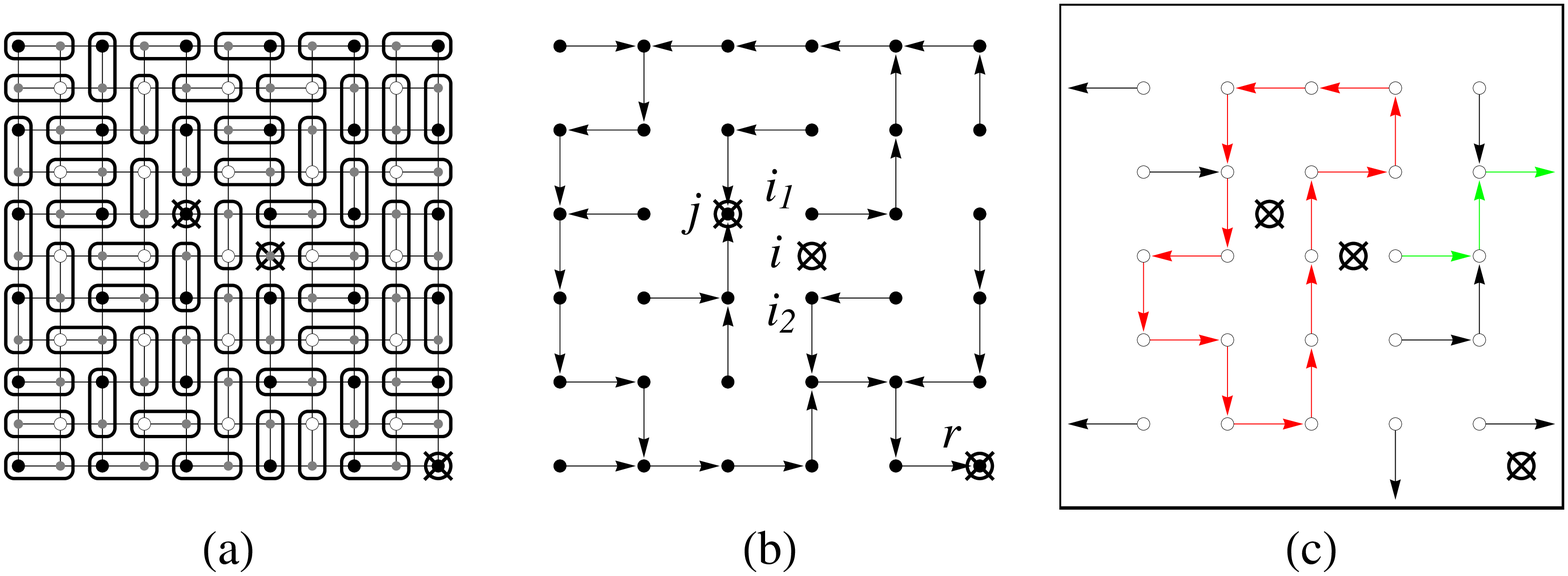}
\vspace{-5mm}
\caption{\label{dim4} The tiling and spanning graphs for case (III).}
\end{figure}

\begin{figure}[!ht]
\includegraphics[width=150mm]{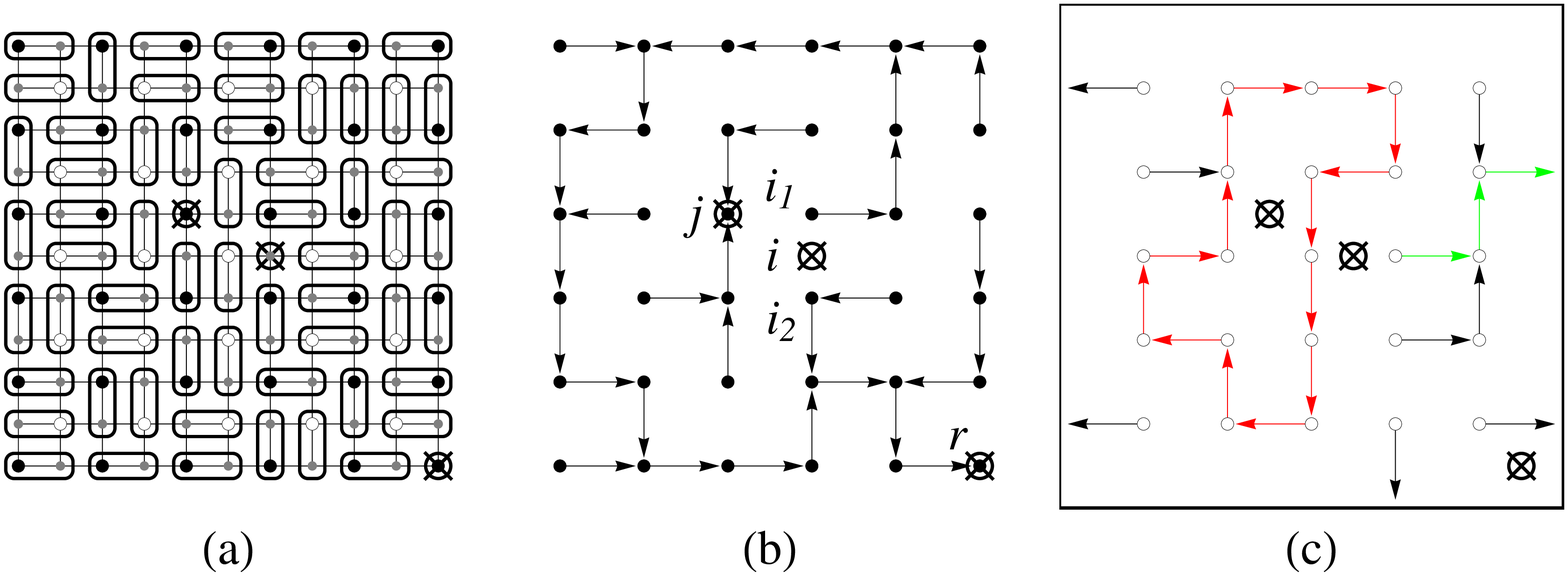}
\vspace{-5mm}
\caption{\label{dim5} The same as previous figure but with the opposite orientation of the loop.}
\end{figure}

\begin{figure}[!ht]
\includegraphics[width=130mm]{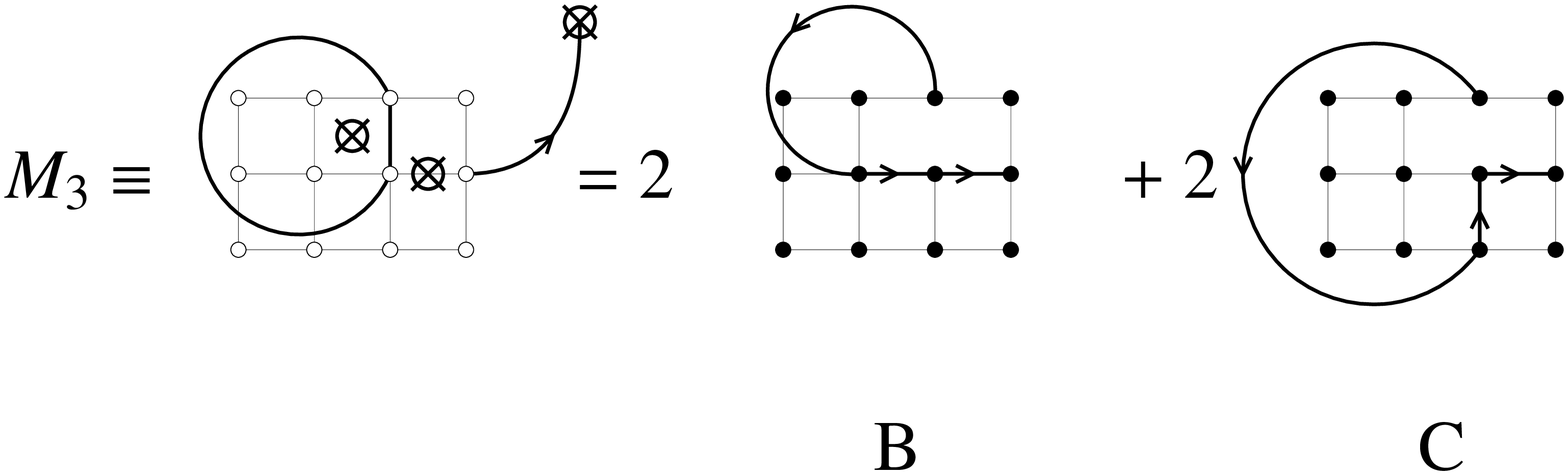}
\vspace{-8mm}
\caption{\label{pred5} Configurations of path for case (III).}
\end{figure}

(III) The sites $i_1$ and $i_2$ belong to the same component, rooted at $r$, and the chemical path between the two loops around $j$, almost completely enclosing the component rooted at $j$, see Fig.\,\ref{dim4}b and Fig.\,\ref{dim5}b. It implies that the dual graph on $\mathcal{W}$ contains one loop around the site $j$ (and only one for the lack of space between $i$ and $j$) . This loop can have the two orientations, as pictured in red in Fig.\,\ref{dim4}c and Fig.\,\ref{dim5}c. Rotating clockwise by $\pi/2$ the red vertical arrow of the loop running in between $i$ and $j$, we obtain a set of paths of type $B$ or $C$, but now on the sublattice $\mathcal{W}$ (Fig.\,\ref{pred5}).
The number of paths must be doubled since there are two equivalent orientations of the arrows along the loop.

Collecting the various contributions from cases (I) to (III), we find how the dimer configurations on $\mathcal{L}'$ is related to the specific classes $A,B,C$ of spanning trees on $\cal B$. Dividing by the total number of dimer configurations on $\cal L$, we obtain that the correlation $P_{mm}$ of two monomers at $i$ and $j$ is expressed in terms of the relative fractions of spanning trees of types $A,B,C$, in the limit of large lattices,
\begin{equation}
P_{mm} = M_1 + M_2 + M_3 = \frac{1}{4\pi} + A + 4 B + 3 C.
\label{MM}
\end{equation}
Following \cite{fist}, the calculation of the correlation is easily carried out. One finds $P_{mm}=\frac{1}{2\pi}$, yielding a first relation for the three unknowns.

\begin{figure}[!ht]
\includegraphics[width=100mm]{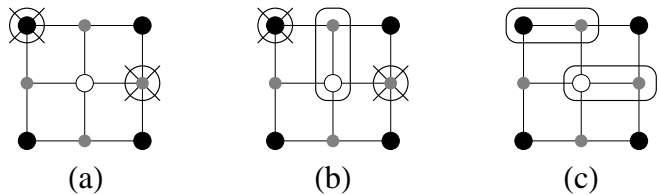}
\vspace{-5mm}
\caption{\label{mon-mon} (a) two monomers; (b) two monomers and one dimer; (c) two dimers equivalent to case (b).}
\end{figure}

In order to write a second relation, we repeat the previous calculation in which, in addition to the two monomers at $i$ and $j$, we force a dimer in between them, like shown in Fig.\,\ref{mon-mon}b. Thus instead of a monomer-monomer correlation, we now consider a monomer-dimer-monomer correlation $P_{mdm}$. All the steps above remain, except that case (I) becomes forbidden (the red loop in Fig.\,\ref{dim2}b would enclose an odd number of sites), and only one orientation of the loop in case (III) is allowed. 
Therefore we can write
\begin{equation}
P_{mdm} = M_2 + \frac{1}{2} M_3 = A + 2 B + C.
\label{MDM}
\end{equation}
The correlation $P_{mdm}$ is equivalent to have two fixed dimers, as shown in Fig.\,\ref{mon-mon}c. From \cite{fist}, we find $P_{mdm} = \frac{1}{8} - \frac{1}{4\pi}$. 

\begin{figure}[!ht]
\includegraphics[width=130mm]{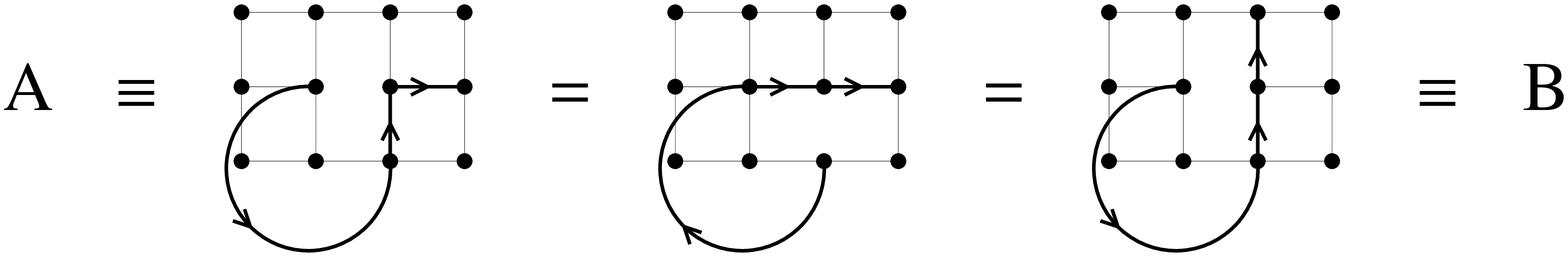}
\vspace{-5mm}
\caption{\label{fig-AB} Three steps proving the equality of $A$ and $B$.}
\end{figure}

Finally we notice the identity $A=B$ as follows from the steps shown on Fig.\,\ref{fig-AB}, where
we reverse the orientation of the loop, move the vertical up arrow to the horizontal right arrow and eventually apply a (diagonal) mirror transformation.

The two equations (\ref{MM}) and (\ref{MDM}) can then be solved, with the result
\begin{eqnarray}
A = &B& = \frac{3}{32} - \frac{1}{4\pi}\,, \label{AB} \\
C &=& \frac{1}{2\pi} - \frac{5}{32}\,. \label{C}
\end{eqnarray}
Plugging these values back in the LERW return probability (\ref{predprob}) yields $P_{\rm ret} = 5/16$, and in turn the ASM mean height $\langle h \rangle = 25/8$.

%%%%%%%%%%%%%%%%%%%%%%%%%%%%%%%%%%%%%%%%%%%%%%%%%%%%%%%%%%%%%%%%%%%%%%%%%%%%%%%%%%%%%%

\section{Perspective}

The present work raises (at least) two natural questions. We have shown how to compute the ASM mean height, a quantity so far thought to have a non-local interpretation in terms of spanning trees, in a purely local way, though specific local arrangements of dimers and monomers. It would be interesting to see whether this technique may help computing 2-site height correlations for heights larger or equal to 2 (since the others are known \cite{madh2,pgpr}). 

The second question is related to the LERW. Beyond the return probability, or passage probability to a nearest neighbour, one could ask for the passage probability to a second nearest neighbour (distance $\sqrt{2}$). From numerical simulations, this latter probability appears to be close to the rational value $2/9$. Applying the present techniques to this case, and perhaps to the next few cases, would bring valuable results.

%%%%%%%%%%%%%%%%%%%%%%%%%%%%%%%%%%%%%%%%%%%%%%%%%%%%%%%%%%%%%%%%%%%%%%%%%%%%%%%%%%%%%%

\begin{acknowledgments}
This work was supported by a Russian RFBR grant No 09-01-00271-a and by the Belgian Interuniversity Attraction Poles Program P6/02,
through the network NOSY (Nonlinear systems, stochastic processes and statistical mechanics).
P.R. is Senior Research Associate of the Belgian National Fund for Scientific Research (FNRS).
\end{acknowledgments}

%%%%%%%%%%%%%%%%%%%%%%%%%%%%%%%%%%%%%%%%%%%%%%%%%%%%%%%%%%%%%%%%%%%%%%%%%%%%%%%%%%%%%%


\begin{thebibliography}{99}

\bibitem{foru} R.H. Fowler and G.S. Rushbrooke, Statistical theory of perfect solutions, {\it Trans. Faraday Soc.}, 1937, {V.33}, 1272-1294.

\bibitem{kast1} P.W. Kasteleyn, Physica \textbf{27}, 1209 (1961).

\bibitem{kast2} P.W. Kasteleyn, J. Math. Phys. \textbf{4}, 287 (1963).

\bibitem{fish} M.E. Fisher, Phys. Rev. \textbf{124}, 1664 (1961).

\bibitem{fite} H.N.V. Temperley and M.E. Fisher, Phil. Mag. \textbf{6}, 1061 (1961).

\bibitem{fist} M.E. Fisher and J. Stephenson, Phys. Rev. \textbf{132}, 1411 (1963).

\bibitem{ken} R. Kenyon, {\it Lectures on dimers}, arXiv:0910.3129 [math.PR].

\bibitem{temp} H.N.V. Temperley, in Combinatorics: Proceedings of the British
Combinatorial Conference, London Math. Soc. Lecture Notes Series V.13, 1974, 202-204.

\bibitem{bupe} R. Burton and R. Pemantle, Ann. Probab. {\bf 21}, 1329 (1993).

\bibitem{bbgj} J. Bouttier, M. Bowick, E. Guitter and M. Jeng, Phys. Rev. E \textbf{76}, 041140 (2007).

\bibitem{ppr} V.S. Poghosyan, V.B. Priezzhev and P. Ruelle, Phys. Rev. E \textbf{77}, 041130 (2008).

\bibitem{law} G.F. Lawler, Duke Math. J. \textbf{47}, 655 (1980).

\bibitem{br} A. Broder, In Symp. Foundations of Computer Sci., IEEE, New York, 442 (1989).

\bibitem{al} D. Aldous, SIAM J. Disc. Math. {\bf 3}, 450 (1990).

\bibitem{pe} R. Pemantle, Ann. Probab. {\bf 19}, 1559 (1991). 

\bibitem{Majumdar} S.N. Majumdar, Phys. Rev. Lett. \textbf{68}, 2329 (1992).

\bibitem{wil} D.B. Wilson, Proceedings of the Twenty-eights Annual ACM Symposium on the Theory of Computing (Philadelphia, PA, 1996), ACM, New York,
296 (1996).

\bibitem{btw} P. Bak, C. Tang and K. Wiesenfeld, Phys. Rev. Lett. \textbf{59}, 381 (1987).

\bibitem{dhar} D. Dhar, Phys. Rev. Lett. \textbf{64} (1990) 1613.

%\bibitem{dhar2} D. Dhar, Phys. Rev. Lett. \textbf{64} (1990) 2837.

\bibitem{madh1} S.N. Majumdar and D. Dhar, Physica \textbf{185}, 129 (1992).

\bibitem{kenyon} R. Kenyon, Ann. Probab. {\bf 29}, 1128 (2001).

\bibitem{iprh} N.Sh. Izmailian, V.B. Priezzhev, P. Ruelle and C.-K. Hu, Phys. Rev. Lett. {\bf 95}, 260602 (2005).

\bibitem{plf} S. Papanikolaou, E. Luijten and E. Fradkin, Phys. Rev. B {\bf 76}, 134514 (2007).

\bibitem{bgpt} J.G. Brankov, S.Y. Grigorev, V.B. Priezzhev and I.Y. Tipunin, J. Stat. Mech. (2008) P11017.

\bibitem{maru} S. Mahieu and P. Ruelle, Phys. Rev. E {\bf 64}, 066130 (2001).

\bibitem{ru} P. Ruelle, Phys. Lett. B \textbf{539}, 172 (2002).

\bibitem{jpr} M. Jeng, G. Piroux and P. Ruelle, J. Stat. Mech. (2006) P10015.

\bibitem{pgpr} V.S. Poghosyan, S.Y. Grigorev, V.B. Priezzhev and P. Ruelle, J. Stat. Mech. (2010) P07025.

\bibitem{schr} O. Schramm, Israel J. Math. {\bf 118}, 221 (2000).

\bibitem{lasw} G.F. Lawler, O. Schramm and W. Werner, Ann. Probab. {\bf 32}, 939 (2004).

\bibitem{popr} V.S. Poghosyan and V.B. Priezzhev, Acta Polytechnica \textbf{51}, 59 (2011).

\bibitem{lepe} L. Levine and Y. Peres, {\it Is the looping constant of the square grid 5/4 ?}, arXiv:1106.2226 [math.PR].

\bibitem{priez} V.B. Priezzhev, J. Stat. Phys. \textbf{74}, 955 (1994).

\bibitem{madh2} S.N. Majumdar and D. Dhar, J. Phys. A: Math. Gen. \textbf{24}, L357 (1991).% L357-L362.

\bibitem{DharGrass} D. Dhar, Physica A \textbf{369}, 29 (2006).

\bibitem{Kenyon} R. Kenyon, Acta Math. \textbf{185}, 239 (2000).

%\bibitem{pgpr1} V.S. Poghosyan, S.Y. Grigorev, V.B. Priezzhev and P. Ruelle, Phys. Lett. B {\bf 659}, 768 (2008).

\bibitem{Fey} A. Fey, L. Levine and D.B. Wilson, Phys. Rev. Lett. {\bf 104}, 145703 (2010).

\end{thebibliography}
\end{document}